# DRAWBACKS AND PROPOSED SOLUTIONS FOR REAL-TIME PROCESSING ON EXISTING STATE-OF-THE-ART LOCALITY SENSITIVE HASHING TECHNIQUES

Omid Jafari[1], Khandker Mushfiqul Islam[2], and Parth Nagarkar[3]

[1, 2, 3] Computer Science Department, New Mexico State University, Las Cruces, USA

## ABSTRACT

*Nearest-neighbor query processing is a fundamental operation for many image retrieval applications. Often, images are stored and represented by high-dimensional vectors that are generated by feature-extraction algorithms. Since tree-based index structures are shown to be ineffective for high dimensional processing due to the well-known "Curse of Dimensionality", approximate nearest neighbor techniques are used for faster query processing. Locality Sensitive Hashing (LSH) is a very popular and efficient approximate nearest neighbor technique that is known for its sublinear query processing complexity and theoretical guarantees. Nowadays, with the emergence of technology, several diverse application domains require real-time high-dimensional data storing and processing capacity. Existing LSH techniques are not suitable to handle real-time data and queries. In this paper, we discuss the challenges and drawbacks of existing LSH techniques for processing real-time high-dimensional image data. Additionally, through experimental analysis, we propose improvements for existing state-of-the-art LSH techniques for efficient processing of high-dimensional image data.*

## 1. INTRODUCTION

Nearest-neighbour queries in high dimensional spaces play a very vital role in a significant part of numerous applications in the image processing domain. While conventional tree-based index structures (such as k-d trees [1], Quadtrees [2], etc.) are a good fit for low-dimensional data, they suffer from the well-known "Curse of Dimensionality" for high dimensional spaces [3]. In high-dimensional spaces, these conventional index structures are often slower than brute-force searches. Looking into approximate results in lieu of exact results is one of the solutions to deal with this problem. Finding approximate results is much faster than finding the exact results in the applications where 100% precision is not necessary. Locality Sensitive Hashing (LSH) [3] is a popular approximate processing technique for processing nearest-neighbour queries in high-dimensional spaces.

Locality Sensitive Hashing (LSH) is based on the idea that adjacent points in the high-dimensional space are also mapped near each other in lower-dimensional spaces. are placed close in the lower dimensional spaces. The process of obtaining these lower-dimensional spaces is through using random projections. Data points are mapped in these random projections in the lower-dimensional space. The goal of the process is to map close points in the original high-dimensional spaces near to each other in the low-dimensional random projections with a high probability. Also, points that are far away in the high-dimensional space should be mapped near each other in the low-dimensional space with a low probability. While querying on this lower-dimensional space, misses and false-positive can occur. Therefore, multiple independently chosen hash projections are used by LSH data structures which are further organized into several hash layers to control the precision of the query processing. Theoretical guarantees on sublinear query time with respect to the dataset size is one of the major advantages of LSH. It makes the use of LSH competent and practical to use.

Image data is often represented by high-dimensional feature descriptors. These high-dimensional descriptors help in efficiently storing and retrieval of these image data. These descriptors are extracted by using popular feature extraction algorithms namely SIFT [4], SURF [5], ORB [6], etc. To search for similar images, high-dimensional features (or feature – depends on which algorithm for the feature extraction is being used) are first extracted from the query image. Then the database that has the features extracted and stored from the images is queried. In particular, similarity search queries are executed for each descriptor in the query image to find the closest matched image. Note that, other domains such as Audio retrieval [7], Video retrieval [8], Computational Biological processing [9], Earth Science data [10], etc., also require high-dimensional data processing.

## 1.1. Paper Organization

The remaining of the paper is organized into several sections. In section 2, we talk about the motivations behind this paper. Section 3 gives a brief overview of several well-known Approximate Nearest Neighbour (ANN) methods. The section is further divided into two subsections, talking about data-dependent techniques and real-time variations. Section 4 specifies the problem is this domain which our method is going to solve. Section 5 explains our proposed method in detail. Next, section 6 presents the experimental evaluation results and an analysis of the results. Finally, section 7 concludes our paper and also presents some suggestions for future work.

## 2. MOTIVATION

With the increase in technological advances, applications in the image retrieval domain face an increasing need for processing fast real-time high dimensional data [11]. Numerous real-world application scenarios vigorously need to find similar objects (also called Nearest Neighbour queries) in real-time. For example, it is an important task to find the near-duplicate images or videos in real-time (for applications which emphasize on video copyright enforcement, content-based video clustering and annotation [12]) on the social Web. In terms of videos, thumbnail images are frequently used to figure out analogous thumbnail images of videos in the present database [11]. Real-time querying of high-dimensional objects is essential for detecting and eliminating image/video duplicates. Also, satellite data, which usually consists of images and remote sensing data, are used to improve decision-making in many earth science applications such as volcanic or earthquake activity monitoring [13], flood management [14], etc. When hazardous situations or natural disasters are occurring, satellites generate important and necessary data more frequently. It is very important for these applications to have real-time indexing and querying support to find comparable situations for the existing data to take future aid decision-making.

## 2.1. Benefits of using Locality Sensitive Hashing

The random hash functions that Locality Sensitive Hashing uses are data-independent, i.e. data characteristics such as data distribution are not needed to generate these random hash functions. On the other hand, data-dependent methods require an analysis of the data during index construction. For static data, this overhead cost is an offline process and hence does not affect the query processing time. However, considering real-time data, the flow of data is a continuous process and this overhead cost becomes an online cost, i.e. the query processing time is affected by the ability of the index structure to index the data. If the indexing is slower, the query processing time will be affected negatively. The indexing process will not add an overhead to the overall processing time if there are no overhead costs, such as analysis of the data distribution. Since LSH uses random hash functions, the generation of these hash functions is a simple process that takes a negligible time. In addition, the generation of these hash functions is not affected by data distribution. Also, these hash functions do not require any change during runtime as newer data are coming in.

While original LSH index structure suffered from large index sizes [15, 16], state-of-the-art LSH techniques [17, 18, 19] have alleviated this issue by using advanced methods such as Collision Counting and Virtual Rehashing. Thus, owing to their small index sizes, and most importantly, lack of any required expensive offline processes, LSH-based techniques are more suitable than other approaches for real-time Approximate Nearest-Neighbour (ANN) processing of high-dimensional image data.

## 3. RELATED WORK

In this section, we give a brief overview of the state-of-the-art techniques in the Approximate Nearest Neighbour (ANN) problem domain in high-dimensional spaces. Since our focus is on Locality Sensitive Hashing (LSH), we focus on LSH and its variants. Also, since our focus is on efficiently solving the ANN problem for real-time environments, we discuss the state-of-the-art techniques with respect to their applicability in real-time environments. Indexing techniques in the ANN problem domain should have these following "good" characteristics [15]: 1) Ability to return results with very low latency, 2) Ability to return results with high accuracy, 3) Index size should be as small as possible and preferably should be linear in the data size, and 4) Practical to use (i.e. the user should not be required to input non-trivial parameters that can drastically change the performance and are specific to each dataset).

## 3.1. Data-Dependent Techniques

Data-dependent hashing techniques [20, 21, 22] provide high accuracy but require offline hash function learning and hash code computation. In addition, these techniques require a high sampling rate for higher accuracy which incurs a high training cost. These methods have been shown to be very slow or infeasible when handling large scale data (when the data points are in billions) [23, 24]. Combined with the additional challenge of handling real-time data, these methods are also inefficient for processing real-time ANN queries.

Similar to data-dependent hashing techniques, Convolutional Neural Networks (CNNs) have become very popular in finding similar media objects [25]. Multi-view alignment hashing (MVAH) [26] learns hash codes with regularized kernel non-negative matrix factorization. It considers both the hidden semantics and joint probability distribution of multiple visual features. To enhance the weights for changed views and concurrently produce the low-dimensional illustration, it incorporates multiple visual features from different views together [26]. By first propagating through the network, [27] encodes incoming images in a faster manner than conventional hashing methods. Afterward, it quantizes the network outputs to binary code representations. The YOLO model [28] operates on a single scale feature map to reframe object detection as a single regression problem. It predicts detections by running a neural network on a new image at 45 frames per second which accomplishes the mean average precision more than twice of other system works in real-time. CNN-based techniques and the above techniques require expensive offline processes such as training and building the neural networks which make them inefficient (and un scalable) for processing real-time ANN queries [25].

### 3.2. Real-time Variants of Locality Sensitive Hashing

While there are several real-time/streaming applications in high-dimensional spaces that leverage LSH, there have been few works [29, 30, 31] that aim at improving LSH for streaming applications. In [29], the authors present a parallel LSH framework that is designed to handle similarity searches on incoming twitter data. The goal of their cache-conscious model is to improve on the creation and updation of the hash tables (which are based on the original LSH design). Similar to other works on Twitter streaming data, the incremental design of the hash tables is designed for social media data where data can expire. In this work, the authors also propose an insert-optimized delta hash table that is periodically merged into the main LSH structure. [30] works on a fast and effective online generation of LSH signatures for streaming data. This work is only applicable to the original LSH design; meanwhile, their goal is to decrease the size of the hash tables when compared to the original work in the presence of streaming data. These works are not optimized for newer LSH designs (that involve Collision counting and Virtual rehashing). [31] introduces a technique called "oLSH" (for online LSH) which improves buffering of incoming data records by storing similar records together on pages. This work does not try to solve the ANN problem in high-dimensional spaces. Instead, they use the "idea" of LSH of storing similar records together on pages for streaming network data and hence call their technique "oLSH".

## 4. PROBLEM SPECIFICATION

Given a multidimensional database D that contains points that belong to a bounded multidimensional space S, the k-NN version of the Approximate Nearest Neighbour problem returns c-approximate results for a given query. Here, c is an approximation ratio greater than 1 and k is the desired number of objects that the user wants to be returned. The goal is to return the correct results with a user-specified success probability, $1-\delta$.

## 5. DRAWBACKS AND PROPOSED SOLUTIONS OF CURRENT STATE-OF-THE-ART TECHNIQUES

In this section, we first present the drawbacks for processing real-time image data of the two existing state-of-the-art techniques, C2LSH [17] and QALSH [18]. As mentioned in Section 2, real-time query processing systems require fast indexing of incoming data and low latency in query execution. The original LSH index structure [3] required many hash layers of several hash functions in order to return high accuracy for the input queries. Additionally, the width of the hash bucket of the hash projections had to be pre-determined before populating the index structure.

### 5.1. Collision Counting Locality Sensitive Hashing (C2LSH)

C2LSH was proposed in [17] that solved these problems by introducing a *collision counting* method that effectively reduced the index sizes by reducing the number of required hash functions. Additionally, it introduced *Virtual Rehashing* that solved the problem of pre-determination of the width of the hash bucket. In order to use these techniques and return correct results with a 1-δ success probability (where δ is a user-specified probability), C2LSH requires *m* hash projections to be built. By using Collision Counting, C2LSH avoids creating multiple hash functions in each layer and instead only requires one hash function per layer, and by using Virtual Rehashing, C2LSH can automatically determine the query radius in order to return top-k results. Additionally, C2LSH can also determine the number of layers required for a dataset (which depends on the cardinality of the dataset) to satisfy a given success probability. While the memory footprint was much lower and the parameter tuning drawback was solved, the accuracy of C2LSH was still not high [18].

**Implementation details**: for each projection, C2LSH stores a bucket number for each dataset point. These values are then copied into a vector and sorted. *It is worth mentioning that sorting can be done on the array itself and copying into a vector and sorting the vector has lower efficiency.* For each projection, a separate binary index file is created. The number of buckets is written to the file as the first 4 bytes. The header size of this file is 4 bytes for the number of buckets and 8 bytes (4 bytes for bucket number and 4 bytes for its offset) for each of the buckets. After writing the header, the algorithm writes the ids belonging to each bucket in the file. As a result, having a bucket number, it is easy to get its offset and then seek to the offset and get all point ids in that bucket. This process is repeated for all projections.

In the query processing phase, the whole dataset file is loaded into an array and parameters are calculated based on the dataset array. The hash functions are also read from the file. Then, the query set and ground truth files are loaded into their respective arrays. Next, the collision counting process (explained below) is done for each desired top K value and each query in the query set. After collision counting, the resultant distances are compared to the ground truth and the ratio is calculated.

In the collision counting process, the hash value (and bucket number) of the query is calculated using the hash functions generated in the indexing phase. Then, based on a page size, which is user input, the initial radius (the number of buckets to process) is determined so that the size of these buckets is not greater than the page size. Based on this radius value, the left and right offsets for all projections are computed. C2LSH then continues processing these hash buckets and increasing the radius as long as the stopping conditions are not met. Also, to process buckets, based on the left and right offsets that were calculated based on the query bucket and radius, C2LSH reads the index files (from the secondary storage) in order to get the point IDs in that range and then increment the collision counts for those points. Each point that has a collision count more than or equal to a collision threshold is considered to be a candidate. Finally, the Euclidean distance is computed for the candidates and false positives are removed in order to return the final results to the user.

**Proposed improvements to C2LSH**: To make this algorithm work in real-time and efficient for streaming applications, everything will need to be stored in memory for faster indexing and retrieval of real-time data. Right now, the collision counting is based on the file offsets and no actual buckets are involved. Basically, in the real-time scenario, after the arrival of each new data, we need to update the indexes and add the incoming data in them. One naïve way of doing this is to recreate the whole indexes as the new data arrives and build them from scratch using the new and previous data. However, since for each indexing phase, files need to be written to and read from the disk, the I/O cost is not efficient and will be a bottle-neck. Also, if a query arrives while the algorithm is still adding a new point to the indexes, the query processing will be further affected negatively. But since these are impractical requirements for many big data applications (where the index size is larger than the given memory size), the other option that should be considered is to store a *delta hash projection* in main memory while the existing projection is stored on the secondary storage. This *delta hash projection* will index the incoming points so that expensive I/Os can be reduced for indexing the incoming points onto the projections on the disk.

The delta hash projection can be implemented using dynamically allocated data structures which can be easily expanded as the new data arrives and is optimized for insertions. However, the drawback of using delta hash projection is that since internally their data is not stored contiguously, they are not optimized for querying. As a result, the delta hash projection eventually needs to be merged with the static data structure residing on the secondary storage. For this purpose, the users can have a threshold that defines the trade-off between insertion speed and querying speed.

## 5.2. Query-Aware Locality Sensitive Hashing (QALSH)

QALSH [18] was introduced that utilized these two novel techniques to improve upon accuracy (at the slight expense of performance). QALSH introduced query-aware hash functions by creating a B+-tree on each random projection and performing incremental range queries until top-k candidates are found. Note that, these hash functions are not created based on the query point, but rather the range queries are more intelligently executed based on the query point. Both of these approaches have one main restriction: as the dataset size grows, more hash functions are required to satisfy the query accuracy guarantee. Hence, as the dataset size grows, new projections need to be created in run-time and all dataset points need to be hashed (and these hash values have to be subsequently sorted) on to this new projection. Additionally, in order to keep the accuracy high, QALSH builds B+-trees on top of each projection, whose creation can be an expensive operation when new projections are created.

**Implementation details**: The hash functions that QALSH uses are also generated randomly like C2LSH. Hash values of all dataset points in each projection are calculated and stored in arrays along with the point IDs. The arrays are sorted in ascending order of the hash values. Based on the page size, each index and leaf node in the B+-tree will have a certain amount of capacity. For page size of 4096 bytes, each leaf node can hold up to 1018 entries (since the point IDs are stored as integers and each integer requires 4 bytes to be stored), and each index node can hold up to 510 entries. Hash values are divided into 1018 sized parts and the lowest hash value in each part is used as the key of the leaf node. The tree is then built in a bottom-up fashion. After building the leaf nodes, each index node holds the keys for 510 leaf nodes. After building the trees for all projections, they're written into binary files. In the binary files, the page size, number of nodes, and several zeroes are written as the header. The header is followed by the data in leaf nodes and index nodes. Since the size of each node was set to be equal to the page size, it's easy to seek to a specific node and read its content. One of the issues with the current implementation of the B+-tree is that the algorithm is not benefiting much from the B+-tree as the number of index nodes is too small compared to the number of leaf nodes. Basically, that means that most of the operation is focused on doing range searches in the leaf nodes, without actually using the benefit of having a hierarchical tree-based structure. For example, for a dataset that contains 1 million 128-dimensional points (that are extracted using the SIFT algorithm [4]), each B+-tree contains 983 leaf nodes and only 2 index nodes.

The collision counting technique is slightly different from C2LSH since QALSH uses B+-trees. In the collision counting phase, the hash values of the query are computed and based on those, the leaf nodes where the query belongs to are found. Moreover, a threshold will be defined as w * radius / 2.0f. This threshold will define the number of range queries to be performed in each radius value. Afterward, the algorithm enters a loop that increases the radius in each iteration, and as a result, the threshold gets increased in each iteration. The algorithm starts with the query node and retrieves the left and right neighbouring nodes based on the threshold. For all IDs in these nodes, a frequency gets incremented at each time. When a frequency of a point reaches the collision threshold, that point will be added to the candidate list. The Euclidean distance is also computed for the candidates, and as soon as the stopping conditions are met, the algorithm breaks out of the loops. *One main performance issue in this algorithm is that the range searches are being done in a bidirectional manner which results in more disk seeks*. Moreover, if the query resides at the end of node X, at radius 1, the algorithm retrieves node X+1 instead of node X. Doing this, the algorithm is skipping the points in node X that might be near the query at radius 1. Also, the usage of the radius is not similar to the logical interpretation of it. The only effect that the radius has in this algorithm is changing the range search threshold. Therefore, a radius value of 2 does not mean that the algorithm will retrieve 2 neighbouring nodes.

**Proposed improvements to QALSH**: QALSH returns more accurate results than C2LSH, mainly due to their usage of the B+-tree to provide query-aware hash functions. This improvement in the result accuracy comes with an overhead of maintaining the B+-trees. For static data, this overhead is worth the improvements in accuracy. But for real-time data, inserting and indexing data into the B+-trees incurs additional overhead. This overhead is dominant especially when the B+-trees are stored on the disk. Note that, every projection in the index structure has a B+-tree built on it. Thus, for each incoming data point, $m$ B+-trees need to be updated. Additionally, similar to the *delta hash functions* introduced in Section 5.1, *delta B+-trees* should be used for efficient handling of real-time data. LSM-trees [29] are a popular indexing paradigm used to handle real-time data. While extending the B+-trees to LSM-trees is a simpler task, extending the concepts of Collision Counting and Virtual Rehashing to LSM-trees is not trivial. Thus, in order to efficiently handle real-time high-dimensional data, modified LSM-trees should be created on top of each of the $m$ projections. LSM-trees involve two components: one tree (called C0) that is stored in the main memory, while another tree (called C1) is stored on the secondary storage. Incoming data is added to C0 (in order to avoid the expensive I/Os of updating a tree on the secondary storage), and when C0 becomes full, C0 is merged with C1 and the process continues. Thus, Collision Counting will need to be updated so that it can run concurrently and efficiently to deal with two B+-trees.

## 6. EXPERIMENTAL EVALUATION

In this section, we present the experimental evaluation of C2LSH and QALSH for real-time scenarios. All experiments are performed on machines with the following configurations: Intel Core i7-6700, 16GB RAM, 2TB HDD, Ubuntu 16.04 OS, and GCC 7.4.0. The reported results are averaged over 50 queries to reduce any bias. The results are showing a simulation of the streaming implementation of C2LSH and QALSH. The codes are written in C++-11 and provided by the authors. Timers are added using the *chrono* library. We use the same settings for both algorithms (c=2, w=2.7191, δ=0.1).

### 6.1. Datasets and Queries

We use several datasets to provide a fair comparison for the two algorithms, C2LSH and QALSH. In particular, we used the following datasets:

- Mnist dataset [33] was created from NIST's Special Database 3 and Special Database 1 which contains binary images of handwritten digits. The resulting dataset has 60,000 examples and 50 dimensions,
- Sift dataset [34] which contains the Sift descriptors extracted from images, and
- Audio dataset [35] which contains the first 192 features extracted using the Marsyas library [36] from the telephone conversations.

These datasets cover both image and audio applications. From these datasets, we generated sub-datasets with different cardinalities to help us with the simulation. Table 1 shows a summary of these datasets. In our simulated streaming scenario, we assume there are already some data points in the system that are already added to the indexes (20,000 for Mnist – 400,000 for Sift – 10,000 for Audio). For the simulation, since some parameters such as the number of projections (and B+-trees) depend on the cardinality of the dataset, we assume the final cardinality is the last cardinality in Table 1, and create the indexes and parameters based on that.

Table 1. Datasets characteristics

| Name  | Dimensions | Cardinalities |
|-------|------------|---------------|
| Mnist | 50         | 20,000 – 30,000 – 40,000 – 50,000 – 60,000 |
| Sift  | 128        | 400,000 – 600,000 – 800,000 – 1,000,000 |
| Audio | 192        | 10,000 – 20,000 – 30,000 – 40,000 – 50,000 |

For the queries, we chose the first 50 points from each dataset as the query set. Since the dataset points are shuffled themselves and LSH does not depend on the order of the points, there was no need to use random queries. Besides, choosing the first 50 points which are common between all cardinalities of the same dataset, guarantees that the comparison is fair.

### 6.2. Evaluation Criteria and Parameters

We evaluate the accuracy of the algorithms (ratio), which is computed based on the outputted distances and ground truth distances. Equation (1) shows how the ratio is calculated. In this equation $o_i$ is the i$^{th}$ object returned by a method, $o_i^*$ is the true i$^{th}$ nearest object, i = 1, 2, 3, …, k, and $\|x, y\|$ denotes the Euclidean distance between two points x and y. We also provide detailed timers to simulate the indexing, query time, and disk I/O size of the streaming scenario. Since we do not change the underlying logic of the algorithm, note that the accuracy of the algorithms does not change (when compared to the original codes).

$$\frac{1}{k}\sum_{i=1}^{k}\frac{\|o_i, q\|}{\|o_i^*, q\|} \quad (1)$$

### 6.3. Discussion of the Performance Results

Although that in many applications, indexing is an offline process and its time is not included in the total processing time; however, in the streaming applications where there is a need to constantly index the new incoming data, the indexing phase of the algorithms should also be considered as important. Therefore, we measured the indexing time on different datasets and different cardinalities. The results of this experiment are shown in Fig. 1. This figure shows the benefits of QALSH indexing method over C2LSH algorithm. The reason behind having a better indexing time in the QALSH algorithm is that QALSH uses an already organized data structure (B+-tree), where it is divided into index and leaf nodes. Furthermore, since B+-trees have been around since a long time ago, they have been fully optimized over the years and their overhead is small. However, C2LSH uses its own custom data structure using the file offsets. Building this custom data structure has a larger overhead compared to QALSH. Another observation from Fig. 1 is that C2LSH is not scalable as the cardinality of the dataset increases. The index structure in C2LSH only contains one header and increasing the data size will make a lot of header data included in this single header structure which makes it inefficient.

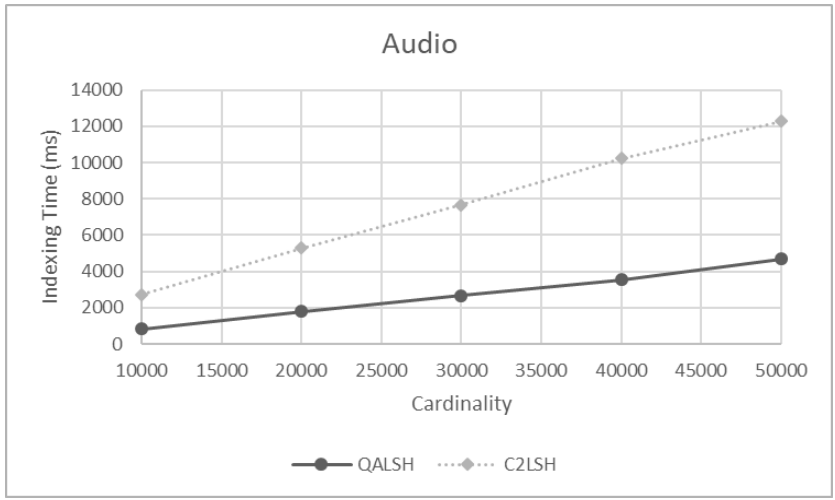

Figure 1.a. Effect of cardinality on Indexing Time in the Audio dataset

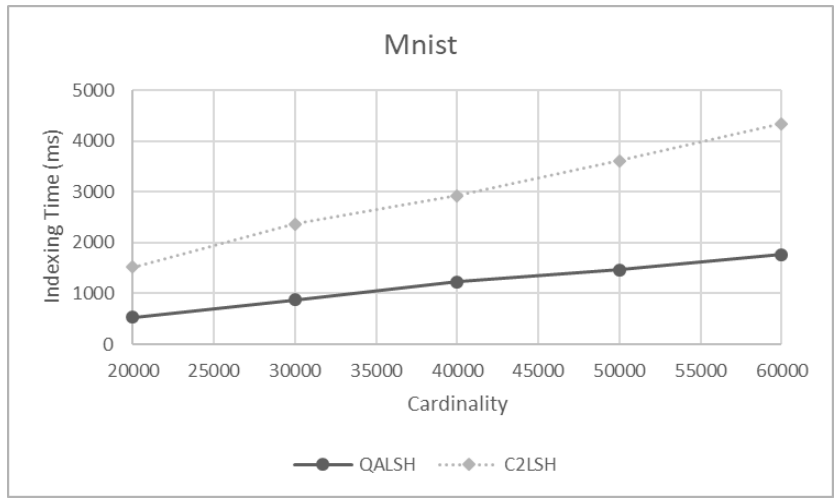

Figure 1.b. Effect of cardinality on Indexing Time in the Mnist dataset

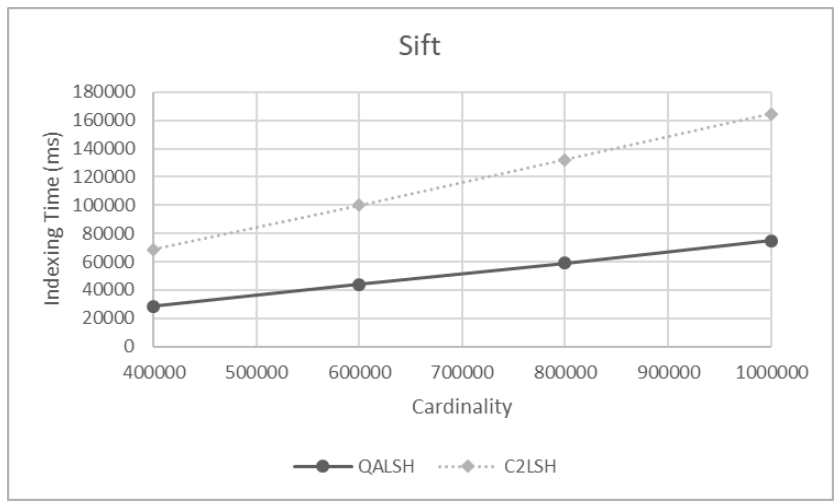

Figure 1.c. Effect of cardinality on Indexing Time in the Sift dataset

We then show the effect of cardinality on the query time for the different datasets in Fig. 2. Query time is mainly the most important factor for users in offline applications. In terms of query time, our analyses showed us that our datasets are not large enough to fully benefit the features of a B+-tree. Therefore, a simple custom structure like C2LSH index structure will serve the users better in the query processing phase. It can also be seen in Fig. 2, C2LSH has a better query time compared to QALSH. For the Sift dataset, which is one of the common datasets in Big Data researches, the query time of QALSH is about 50% times more than the query time in C2LSH. The figure also shows us that C2LSH is more scalable as the cardinality increases. Talking about scalability, a large B+-tree will have a lot of vertical and horizontal traversals, which will make the querying slow, and that is the reason why QALSH is not having good scalability in terms of query processing.

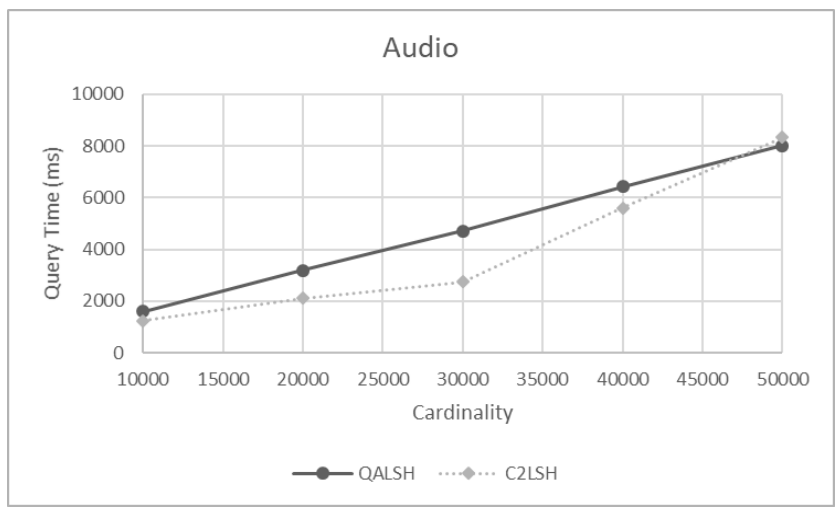

Figure 2.a. Effect of cardinality on Query Time in the Audio dataset

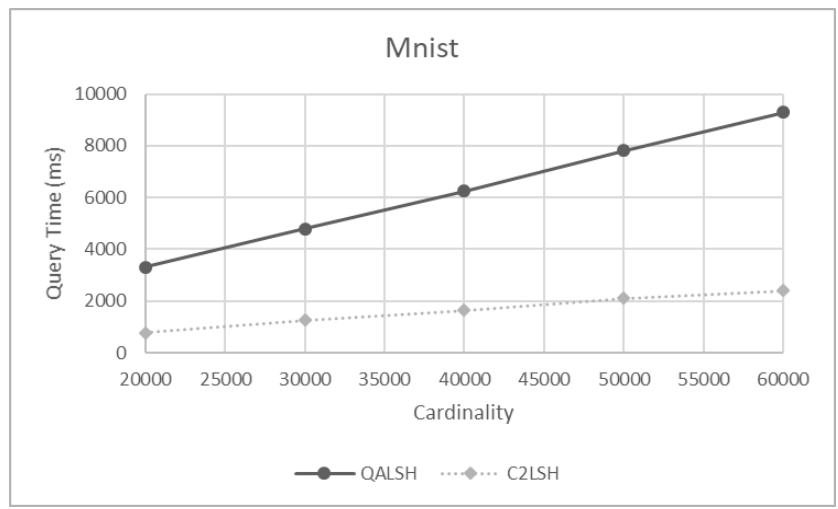

Figure 2.b. Effect of cardinality on Query Time in the Mnist dataset

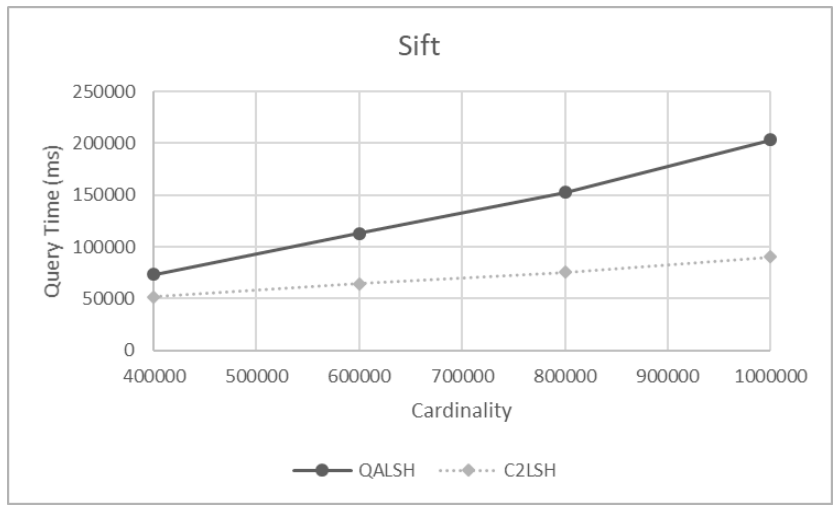

Figure 2.c. Effect of cardinality on Query Time in the Sift dataset

The main purpose of the ANN methods is to provide users with the most accurate results in the shortest time possible. Therefore, a lower ratio will help users have the closest results to the exact nearest neighbors. Fig. 3 shows the ratio for different cardinalities of the datasets. Since a ratio value of one shows the best accuracy and considered to be the baseline, all of these ratio values are near one and considered accurate. Results also show that as the cardinality increases, QALSH will have a better ratio value compared to C2LSH.

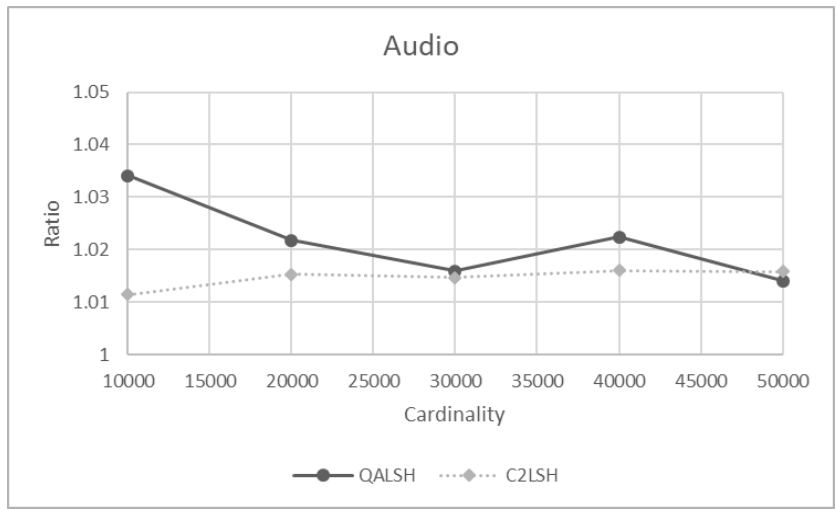

Figure 3.a. Effect of cardinality on Ratio in the Audio dataset

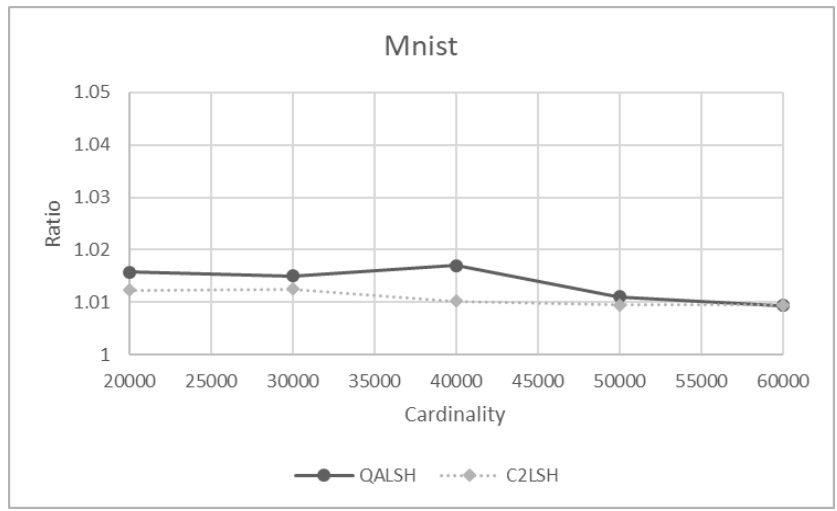

Figure 3.b.  Effect of cardinality on Ratio in the Mnist dataset

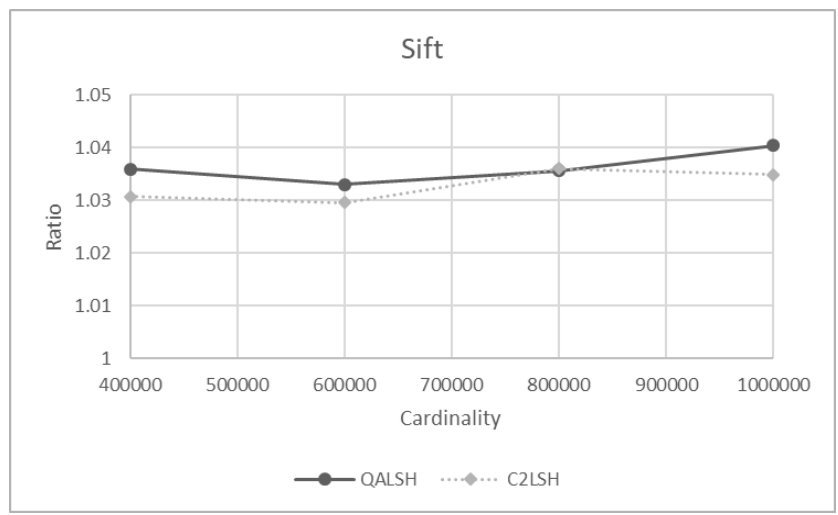

Figure 3.c.  Effect of cardinality on Ratio in the Sift dataset

## 7. CONCLUSIONS

In this paper, we present the challenges of real-time processing of image data. Through extensive analysis, we also discuss the drawbacks of existing state-of-the-art Locality Sensitive Hashing techniques. These techniques suffer from several drawbacks that make them unsuitable for real-time processing of high-dimensional image data. Additionally, we present our algorithmic and implementation analysis of the existing state-of-the-art locality sensitive hashing-based algorithms. Further, we proposed improvements over these techniques that can improve the performance of these techniques in real-time environments. Our experimental analysis confirms our algorithmic and implementation analysis.

## 8. FUTURE WORK

In the future, we are planning to implement our proposed technique and also apply it to other ANN techniques. This will help us compare and observe the benefits of our proposed technique over several ANN techniques. We also plan on comparing with the existing real-time ANN techniques which are not based on Locality Sensitive Hashing. In the paper, we mentioned that there are several trade-offs between insertions and querying and also between accuracy and performance. Our future research will also be focused on analyzing these trade-offs and finding the optimal configurations for given scenarios.

## AUTHORS

**Omid Jafari** is a Ph.D. student in the Computer Science department at New Mexico State University. He got his master's degree in Software Engineering from Azad University of Mashhad, Iran in 2017. He is currently researching in the area of Big Data Management; particularly, high-dimensional data and query processing.

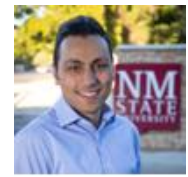

**Khandker Mushfiqul** Islam received a Bachelor's in Science degree from the School of Engineering and Computer Science, BRAC University, Dhaka, Bangladesh, in 2017. He is currently pursuing a Ph.D. degree in Computer Science at New Mexico State University. His research interests are in Big Data management with a focus on query optimization.

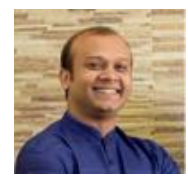

**Parth Nagarkar** is an Assistant Professor in the Computer Science Department at New Mexico State University. He received his Ph.D. from Arizona State in 2017. His research interests are in the broad area of big data management. In particular, he is interested in building scalable and efficient index structures for processing large and high-dimensional data.

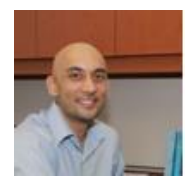